\documentclass[onecolumn
reprint,
preprint,
 amsmath,amssymb,
 aps,
prd
]{revtex4-2}

\usepackage{hyperref}
\usepackage{graphicx}
\usepackage{dcolumn}
\usepackage{bm}
\usepackage{float}
\usepackage{subcaption}
 \usepackage{xcolor}

\usepackage[mathlines]{lineno}

\begin{document}


\title{Analytic forms for the $e^+e^-$ annihilation cross sections around a resonance including initial state radiation}

\author{Baoxin Liu}
\author{Zhenyu Zhang}
\email{zhenyuzhang@whu.edu.cn}
\author{Xiang Zhou}
 \email{xiangzhou@whu.edu.cn}
\affiliation{%
 Hubei Nuclear Solid Physics Key Laboratory, School of Physics and Technology, Wuhan University, Wuhan, Hubei 430072, People's Republic of China
}%

\date{\today}

\begin{abstract}
For the first time, the exact analytic integration of the multiplication of the Kuraev-Fadin radiative function and the Born-level cross section has been achieved for $e^+e^-$ annihilation around a narrow resonance. The analytic result is in a perfect agreement with the one obtained through direct numerical integration. The extraction of resonance parameters is demonstrated through the unfolding of the initial state radiation effect. The analytic integration  significantly accelerates the regression process by more than 160 times compared to numerical integration. Moreover, it facilitates the determination of experimental measurements for physical parameters, such as the two-dimensional confidence region of the branching fraction and the relative phase between the strong and electromagnetic amplitudes, within just one hour instead of one week typically required for numerical integration.

\begin{description}
\item[Keywords] Initial state radiation, cross section, branching fraction, analytic form.

\end{description}
\end{abstract}

\maketitle

\section{Introduction}
The $e^+e^-$ annihilation experiments provide cleaner experimental environments for quarkonium decays than $p\bar p$ and hadron decays experiments. The measured quarkonium decay widths, or the corresponding branching fractions, can serve as inputs in phenomenological models or be used for comparison with theoretical predictions to test our understanding of quantum chromodynamics (QCD)~\cite{Brambilla2011}. There is an unavoidable background from the continuum process, which directly produces the final state via $e^+e^-$ annihilation, i.e., $e^+e^-\to\gamma^*\to f$~\cite{Wang2004}.  It has been shown that the cross section from the interference term will lead to imprecise branching fractions for the broad resonance above open heavy flavor threshold, such as $\psi(3770)$~\cite{Wang2006}. 

A novel study shows that the ratio of the cross section from the interference term with respect to the resonance is surprisingly large compared to the precision of the current experiments even for the narrow resonances below the open heavy flavor threshold, such as $J/\psi$ and $\psi(3686)$~\cite{Guo2022}. Therefore, to achieve precision measurements of the quarkonium decay widths or the corresponding branching fractions, the cross sections around the resonance of at least three energies must be measured~\cite{Guo2022}.

While the Born cross section of the final state is of interest, it is the experimentally observed cross section data that are measured which contain the initial state radiation (ISR) effect~\cite{Dong2018,Gribanov2021}. ISR is an essential quantum electrodynamics (QED) correction for the precise measurement of cross sections in $e^+e^-$ annihilations. In the ISR process, the electron or positron emits one or more photons before annihilation, thereby reducing the center-of-mass (c.m.) energy. One well-known method to take into account the ISR effect is the structure function method which is an integral transformation, i.e. convolution, of the kernel cross section by the radiative function~\cite{E1985}. Therefore, obtaining the Born cross section is mainly an unfolding process of the ISR effect.
 
One commonly unfolding method involves employing an iterative procedure to obtain the discrete Born cross sections from the experimentally observed cross section data and by using the Monte Carlo method to obtain the uncertainties~\cite{Dong2018,Sun2021}. Recently, a new method, named as the naive method, can obtain both the discrete Born cross sections and the uncertainties simultaneously by solving numerical integral equations~\cite{Gribanov2021}. The advantage for the above methods is that their results are convenient to compare with results obtained in other experiments. With the obtained discrete Born cross sections, the parameters, such as decay width or branching fraction can be determined by fitting with a certain model.

However, the above methods cannot apply to the $e^+e^-$ annihilation processes around a narrow resonance, such as $J/\psi$ and $\Upsilon(1S)$, because the Born cross section varies sharply and the c.m. energy spread is much greater than the resonance width and the distances between the c.m. energy points of data around the resonance. One of the most frequently used methods is to fit the experimentally observed cross section data by the Born cross sections specified in a certain model with the effects of both ISR and c.m. energy spread~\cite{YaDi2019}. The effect of the c.m. energy spread is taken into account as an additional Gaussian convolution. The integral method is straightforward but time-consuming, as it requires numerical integrations in the regression iterations.

The integral method will be significantly accelerated if the analytic form of the ISR-corrected cross sections is obtained~\cite{Yanan2023}. It is well known that the radiative function proposed by Kuraev and Fadin (KF) formulism consists of an exponentiated part and a finite-order leading-logarithmic part, accounting for soft multiphoton emission and hard collinear bremsstrahlung, respectively~\cite{E1985}. In 1987, R. N. Cahn was the first to derive the analytic form with the exponentiated part in the KF radiative function~\cite{Cahn1987}. The approximated analytic form, including the leading-logarithmic parts in the KF radiative function and the upper limit correction by the exponential expansions, has been developed in subsequent works~\cite{chen1990,zhou2017,YaDi2019,Yanan2023}. However, it is shown below that for the ISR-corrected cross sections around $J/\psi$, the accuracy of the approximated analytic form is within a few percent and propagates in the same order to the estimated value by regression for the hadronic branching fraction. In a sense, the approximated analytic form introduces an imperceptible yet non-negligible systematic uncertainty, which is comparable to the statistical or even the total systematic uncertainty for experiments. 

In this paper, we provide the formalism of the ISR corrected Born cross sections and validate the precision of the exact analytic integrations in Sections~\ref{ISR correction to Born cross sections}, ~\ref{Vacuum polarization effect} and ~\ref{beam energy spread effect} introduce the formalism that incorporate the vacuum polarization effect and the c.m. energy spread effect, respectively, demonstrating the corresponding precision. Section~\ref{Fit results for the toy MC cross section data} presents the fit performance tests of the toy Monte Carlo (MC) samples in the vicinity of $J/\psi$. Section~\ref{Discussions and Conclusions} gives discussions and conclusions. Appendix~\ref{Appendix A} gives the expressions of the Born cross sections around a resonance. Appendix~\ref{AppendixB} provides the corresponding exact analytic integrations of the ISR-corrected cross sections incorporating the KF radiative function, referred as the KF analytic forms.

\section{ISR corrected Born cross sections}\label{ISR correction to Born cross sections}
In the vicinity of a resonance, the amplitude
$A_\mathrm{tot.}^f$ of the final state $f$ in $e^+e^-$ colliders is the coherent sum of
both resonance $A_R^f$ and continuum amplitudes $A_C^f$. The Born
cross section can be written as~\cite{Guo2022}
\begin{eqnarray}
\sigma_\mathrm{Born}^f(W)&\propto&\left|A_\mathrm{tot.}^f(W)\right|^2=\left|A_C^f(W)\nonumber\right.\\&&\left.+A_R^f(W)e^{i\phi}\right|^2,
\end{eqnarray}
where $\phi$ is the relative phase between the continuum
amplitude $A_c^f$ and the resonance amplitude $A_R^f$. Therefore,
the Born cross section is a sum of three parts as
\begin{equation}
\sigma_\mathrm{Born}^f(W) = \sigma_{B_C}^f(W)+\sigma_{B_R}^f(W)+\sigma_{B_I}^f(W),
\end{equation}
where $\sigma_{B_C}^f(W)\propto \left|A_C^f(W)\right|^2$,
$\sigma_{B_R}^f(W)\propto \left|A_R^f(W)\right|^2$ and
$\sigma^f_{B_I}\propto 2\Re\{ A_c^f(W)A_R^f(W)\}$ denote the Born cross
sections from continuum, resonance and interference contributions,
respectively.

When the kernel cross section is the Born cross section, the ISR corrected cross section $\sigma_\mathrm{ISR}^f$ is an integral of the Born cross section $\sigma_\mathrm{Born}^f$ times the radiation function~\cite{Guo2022}, that is
\begin{eqnarray}\label{sigmaISR}
    \sigma_\mathrm{ISR}^f(W) &=& \int_0^{1-\left(\frac{W_\mathrm{min}}{W}\right)^2}\mathrm{d}xF(x,W)\nonumber\\&&\times \sigma_\mathrm{Born}^f(W\sqrt{1-x}),
\end{eqnarray}
where $W$ is the c.m. energy of $e^+e^-$ annihilation and $W_{\text{min}}$ is the threshold energy equaled to the invariant mass of the final states or the experimental cutoff energy. The KF radiative function $F(x,W)$ has the form~\cite{E1985}
\begin{eqnarray}
F(x, W)& = & \beta x^{\beta-1}(1+\delta)-\beta\left(1-\frac{x}{2}\right)+\frac{1}{8} \beta^2\left[4(2-x) \right. \nonumber\\
& & \left.\times\log \frac{1}{x}-\frac{1+3(1-x)^2}{x} \log (1-x)\right. \nonumber\\
& & \left.-6+x\right],
\end{eqnarray}
with
$\delta = \frac{3}{4}\beta+\frac{\alpha}{\pi}\left(\frac{\pi^2}{3}-\frac{1}{2}\right) + \beta^2\left(\frac{9}{32}-\frac{\pi^2}{12}\right)$ and $\beta=\frac{2\alpha}{\pi}\left(2\log\frac{W}{m_e}-1\right)$.

The ISR corrected cross section integral $\sigma_\mathrm{ISR}^f$ is a sum of three parts, which is
\begin{equation}
    \sigma_\mathrm{ISR}^f(W) = \sigma_C^f(W) + \sigma_R^f(W) + \sigma_I^f(W),
\end{equation}
where
\begin{equation}
    \sigma_C^f(W) \equiv \int_0^{1-\left(\frac{W_\mathrm{min}}{W}\right)^2}\mathrm{d}xF(x,W)\sigma_{B_C}^f(W\sqrt{1-x}),
\end{equation}
\begin{equation}
    \sigma_R^f(W) \equiv \int_0^{1-\left(\frac{W_\mathrm{min}}{W}\right)^2}\mathrm{d}xF(x,W)\sigma_{B_R}^f(W\sqrt{1-x}),
\end{equation}
and
\begin{equation}
    \sigma_I^f(W) \equiv \int_0^{1-\left(\frac{W_\mathrm{min}}{W}\right)^2}\mathrm{d}xF(x,W)\sigma_{B_I}^f(W\sqrt{1-x}),
\end{equation}
are the integrals of the continuum, resonance and interference contributions, respectively. 

For the process $e^+e^-\to\mu^+\mu^-$ , in the vicinity of $J/\psi$, the Born cross section is~\cite{YaDi2019}
\begin{eqnarray}\label{eq:Bornmumu}
\sigma_\mathrm{Born}^{\mu^+\mu^-}(W)&=&\frac{4 \pi \alpha^2}{3 W^2}\left|1+\frac{W^2}{M} \right. \nonumber\\
& & \left.\times\frac{3 \sqrt{\Gamma_{e e} \Gamma_{\mu \mu}}}{\alpha\left(W^2-M^2+i M \Gamma\right)} e^{i \phi}\right|^2,
\end{eqnarray}
where $\alpha$ is the fine structure constant, $M$ and $\Gamma$ are the mass and total decay width of $J/\psi$, $\Gamma_{ee}$ and $\Gamma_{\mu\mu}$ are the decay widths of $J/\psi\to e^+e^-$ and $\mu^+\mu^-$, respectively.  For the processes $e^+e^-\to 2(\pi^+\pi^-)\pi^0$ and $\eta\pi^+\pi^-$ with
\(\eta\to\pi^+\pi^-\pi^0\) in the vicinity of $J/\psi$, the Born cross section can be written in a
general form as~\cite{YaDi2019}
\begin{eqnarray}\label{eq:Born5pi}
\sigma_\mathrm{Born}^{5\pi}(W)  &=&\left(\frac{\mathcal{A}}{W^2}\right)^2 \frac{4 \pi \alpha^2}{3W^2} \left|1+\right. \nonumber\\
& & \left.\frac{3 W^2 \sqrt{\Gamma_{e e} \Gamma_{\mu \mu}} \mathcal{C}_1 e^{i \phi}\left(1+\mathcal{C}_2 e^{i \Phi}\right)}{\alpha M\left(W^2-M^2+i M \Gamma\right)}\right|^2,
\end{eqnarray}
where $\frac{\mathcal{A}}{W^2}$ is the form factor, $\mathcal{C}_1$ and $\mathcal{C}_2$ are the ratios of amplitudes, and $\Phi$ is the phase between the strong and electromagnetic decays from $J/\psi$. For the process of $e^+e^-\to 2(\pi^+\pi^-)\pi^0$, $\mathcal{C}_1$ and $\phi$ are assumed to be 1 and 0, respectively~\cite{YaDi2019}. 

It can be derived for the cross sections from continuum, resonance and interference contributions. For example, the Born cross section from the continuum contribution of $e^+e^-\to\mu^+\mu^-$ is
\begin{equation}
\sigma_{B_C}^{\mu^+\mu^-}(W) = \frac{4\pi\alpha^2}{3W^2}. \\ 
\end{equation}

The Born cross sections are composed by rational functions, while the KF radiative function contains not only exponential functions but also logarithmic functions, such as $\log x$, which makes the ISR corrected cross section integral an integrable singularity at the lower limit $x=0$ in Eq.~\eqref{sigmaISR}. Fortunately, the integral still has an analytic form. For example, the analytic form for the improper integral of $\log\frac{1}{x}$ times $\sigma_{B_C}^{\mu^+\mu^-}(W)$ is
\begin{eqnarray}
\mathcal{I}(W) &=& \int_0^{1-\left(\frac{W_\mathrm{min}}{W}\right)^2}\mathrm{d}x\beta^2\log\frac{1}{x}\sigma_{B_C}^{\mu^+\mu^-}(W\sqrt{1-x})\nonumber\\
&=& \int_0^{1-\left(\frac{W_\mathrm{min}}{W}\right)^2}\mathrm{d}x\beta^2\log\frac{1}{x}\frac{4\pi\alpha^2}{3W^2(1-x)}\nonumber\\
&=& -\beta^2\frac{4\pi\alpha^2}{3W^2}\int_0^{1-\left(\frac{W_\mathrm{min}}{W}\right)^2}\mathrm{d}x\frac{\log x}{1-x} \nonumber\\
&=& -\beta^2\frac{4\pi\alpha^2}{3W^2}\left[\mathrm{Li}_2\left(\frac{W_\mathrm{min}^2}{W^2}\right)-\frac{\pi^2}{6}\right],
\end{eqnarray}
where \(\mathrm{Li}_2(z)\) is Spence's function which is defined as
\begin{eqnarray}
\mathrm{Li}_2(z) &\equiv&-\int_0^{z} \frac{\log (1-x)}{x} \text{d} x\nonumber\\
 &=& \int_1^{1-z} \frac{\log (x)}{1-x} \text{d} x.
\end{eqnarray}

Since the  analytic forms are tedious, we put them into Appendix~\ref{AppendixB}.

\begin{figure*}[!htp]
    \centering
    \includegraphics[width=17.2cm]{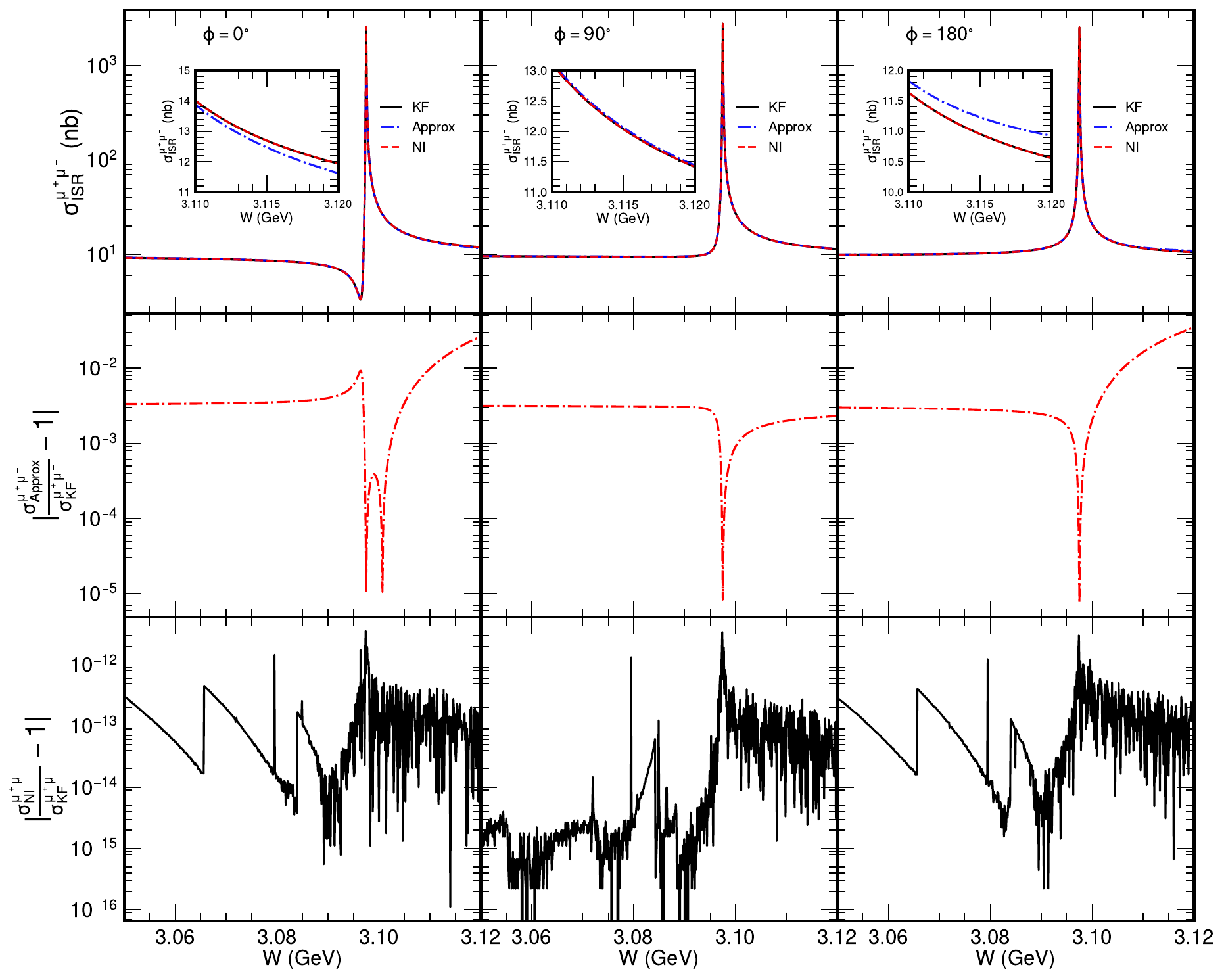}
    \captionsetup{
   justification=raggedright,
   singlelinecheck=false,
   labelsep=period
}
   \caption{\label{fig:observed} The comparisons of ISR corrected cross sections for $e^+e^-\to\mu^+\mu^-$ at $\phi=0^{\circ}$ at $\phi=0^\circ$, 90$^\circ$ and 180$^\circ$. The Born cross section parameters are set to be Particle Data Group (PDG) world averages~\cite{PDG} and $W_\mathrm{min}=2.0$ GeV. In the top row, black lines represent KF method results, blue dashed lines represent approximated analytic form results and red dotted lines represent NI method results. In the middle row, red dashed lines represent the results of $\left|\frac{\sigma^{\mu^+\mu^-}_{\text{Approx}}}{\sigma^{\mu^+\mu^-}_{\text{KF}}}-1\right|$. In the bottom row, black lines represent the results of $\left|\frac{\sigma^{\mu^+\mu^-}_{\text{NI}}}{\sigma^{\mu^+\mu^-}_{\text{KF}}}-1\right|$.}
\end{figure*}

In Fig.~\ref{fig:observed}, we compare the ISR corrected cross section $\sigma^{\mu^+\mu^-}_{\mathrm{NI}}$ obtained through the numerical integration (NI) with the KF analytic form $\sigma^{\mu^+\mu^-}_{\mathrm{KF}}$, and the approximated analytic form $\sigma^{\mu^+\mu^-}_{\mathrm{Approx}}$ for the $e^+e^-\to\mu^+\mu^-$ process around the $J/\psi$ resonance. The  relative deviation between $\sigma^{\mu^+\mu^-}_{\mathrm{KF}}$ and $\sigma^{\mu^+\mu^-}_{\mathrm{NI}}$ is less than $10^{-10}$, coinciding with the numerical integration's error tolerance, thereby affirming the precision of our analytic form. Moreover, the relative deviation between $\sigma_{\text{KF}}^{\mu^+\mu^-}$ and $\sigma_{\text{Approx}}^{\mu^+\mu^-}$ reaches approximately 0.3\% at the incoherent phase $\phi=90^\circ$ where $\sigma_{B_I}^{\mu^+\mu^-}$ and then $\sigma_I^{\mu^+\mu^-}$ vanish. Therefore, the precision is about 0.3\% for the sum of the approximated analytic forms $\sigma_C^{\mu^+\mu^-}$ and $\sigma_R^{\mu^+\mu^-}$ where the precision of the approximated analytic form $\sigma_R^{\mu^+\mu^-}$ is only 0.1\%~\cite{chen1990}. Conversely, at $\phi=0^\circ$ or $180^\circ$, the  relative deviation between $\sigma_{\text{KF}}^{\mu^+\mu^-}$ and $\sigma_{\text{Approx}}^{\mu^+\mu^-}$ is about 3\%. It is evident that the  deviation that comes from $\sigma_I^{\mu^+\mu^-}$ is 3\% in the approximated analytic form. 

\section{Vacuum polarization effect}\label{Vacuum polarization effect}
When the vacuum polarization (VP) effect is considered, the Born cross section in the vicinity of a resonance is corrected as~\cite{Anashin2012,zhou2017}
\begin{eqnarray}
\sigma_{\mathrm{Born-VP}}^f(W) &=& \sigma_{B_C^\mathrm{VP}}^f(W) + \sigma_{B_R^\mathrm{VP}}^f(W) + \sigma_{B_I^\mathrm{VP}}^f(W) \nonumber\\
&=& \frac{\sigma_{B_C}^f(W)}{\left|1-\Pi_0(W)\right|^2} + \sigma_{B_R}^f(W)\nonumber\\ &&+ \frac{\sigma_{B_I}^f(W)}{\left|1-\Pi_0(W)\right|},\label{eq:VP}
\end{eqnarray}
where $\Pi_0(W)$ is the nonresonant VP factor~\cite{Anashin2012}.

The ISR corrected cross section with VP effect is also a sum of three parts which is
\begin{eqnarray}
\sigma_{\mathrm{ISR-VP}}^f(W) &=& \sigma_{C-\mathrm{VP}}^f(W)+\sigma_{R-\mathrm{VP}}^f(W)\nonumber
\\&&+\sigma_{I-\mathrm{VP}}^f(W),
\end{eqnarray}
where 
\begin{eqnarray}
\sigma_{R-\mathrm{VP}}^f(W) \equiv \sigma_{R}^f(W)&=&\int_0^{1-\left(\frac{W_\mathrm{min}}{W}\right)^2}\mathrm{d}xF(x,W)\nonumber
\\&&\times\sigma_{B_R}^f(W\sqrt{1-x}),\label{eq:VR}
\end{eqnarray}
\begin{eqnarray}
\sigma_{C-\mathrm{VP}}^f(W) &\equiv& \int_0^{1-\left(\frac{W_\mathrm{min}}{W}\right)^2}\mathrm{d}xF(x,W)\nonumber
\\&&\times\frac{\sigma_{B_C}^f(W\sqrt{1-x})}{\left|1-\Pi_0(W\sqrt{1-x})\right|^2},\label{eq:VC}\\
&\simeq& \frac{\sigma_C^f(W)}{\left|1-\Pi_0(W)\right|^2},\label{eq:VC-VP}
\end{eqnarray}
and
\begin{eqnarray}
\sigma_{I-\mathrm{VP}}^f(W) &\equiv& \int_0^{1-\left(\frac{W_\mathrm{min}}{W}\right)^2}\mathrm{d}xF(x,W)\nonumber
\\&&\times\frac{\sigma_{B_I}^f(W\sqrt{1-x})}{\left|1-\Pi_0(W\sqrt{1-x})\right|},\label{eq:VI}\\
&\simeq& \frac{\sigma_I^f(W)}{\left|1-\Pi_0(W)\right|},\label{eq:VI-VP}
\end{eqnarray}
are integrals of the resonance, continuum, and interference contributions, respectively. Approximations are used in Eqs.~\eqref{eq:VC-VP} and~\eqref{eq:VI-VP} compared to Eqs.~\eqref{eq:VC} and~\eqref{eq:VI} because the nonresonance VP factors $\Pi_0(W)$ are numerically obtained~\cite{VP2019}. 

\begin{figure*}[!htp]
    \centering
    \includegraphics[width=17.2cm]{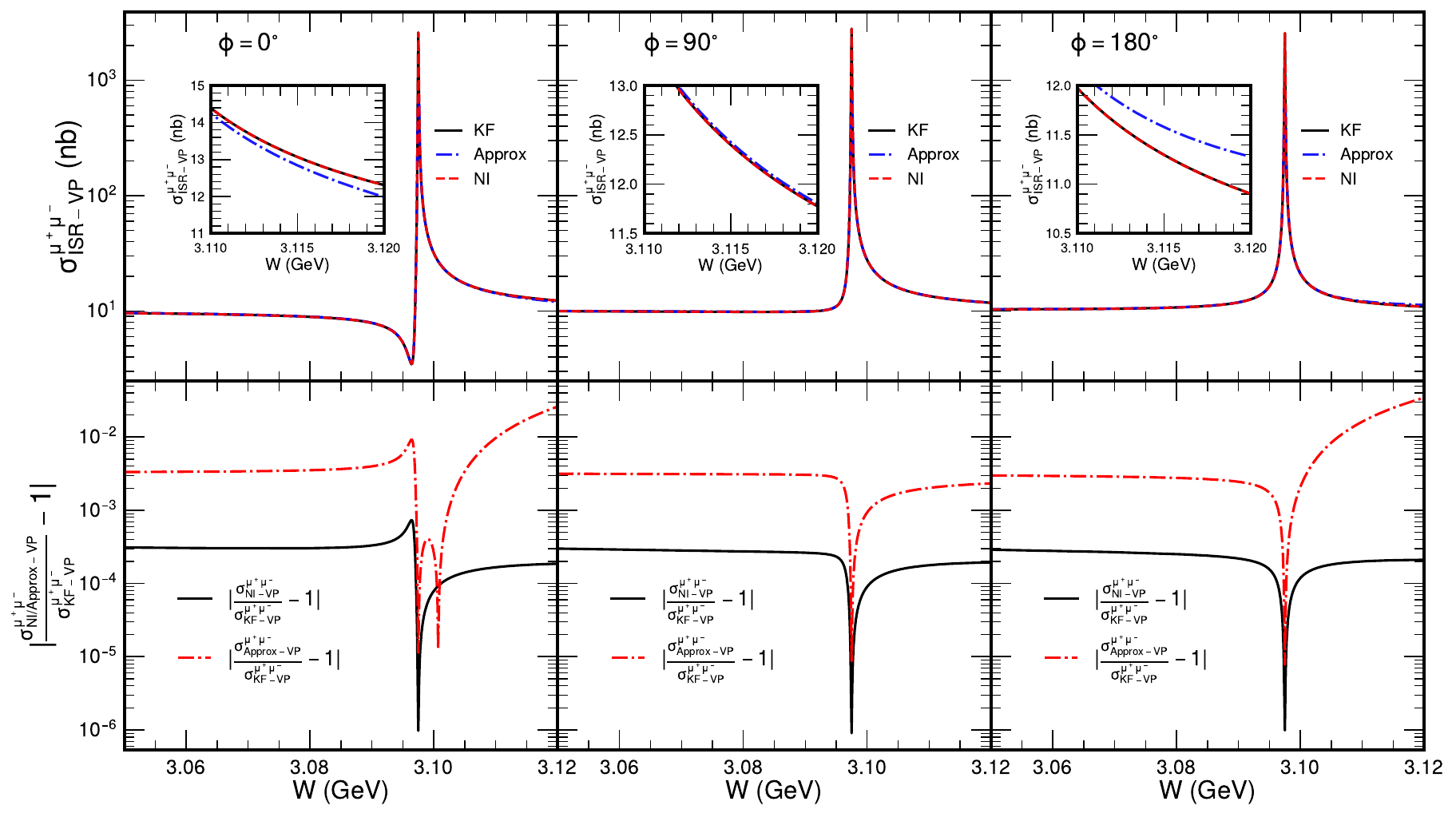}
        \captionsetup{
   justification=raggedright,
   singlelinecheck=false,
   labelsep=period
}
    \caption{\label{fig:visu} The comparisons of ISR corrected cross sections with VP effect for $e^+e^-\to\mu^+\mu^-$ at $\phi=0^{\circ}$, $90^{\circ}$ and $180^{\circ}$. The Born cross section parameters are set to be PDG world averages~\cite{PDG} and $W_\mathrm{min}=2.0$ GeV.  In the top row, black lines represent KF method results, blue dashed lines represent approximated analytic form results and red dotted lines represent NI method results. In the bottom row, red dashed lines represent the results of  $\left|\frac{\sigma^{\mu^+\mu^-}_{\text{Approx-VP}}}{\sigma^{\mu^+\mu^-}_{\text{KF-VP}}}-1\right|$ and  black lines represent the results of $\left|\frac{\sigma^{\mu^+\mu^-}_{\text{NI-VP}}}{\sigma^{\mu^+\mu^-}_{\text{KF-VP}}}-1\right|$. }
\end{figure*}

In Fig.~\ref{fig:visu}, we compare the ISR corrected cross section including VP effect $\sigma_\mathrm{NI-VP}^{\mu^+\mu^-}$ obtained through NI with $\sigma_\mathrm{KF-VP}^{\mu^+\mu^-}$ obtained using the KF analytic form and $\sigma_\mathrm{Approx-VP}^{\mu^+\mu^-}$ obtained using the approximated analytic form. The relative deviation between $\sigma_\mathrm{NI-VP}^{\mu^+\mu^-}$ and $\sigma_\mathrm{KF-VP}^{\mu^+\mu^-}$ is at the order of $10^{-4}$, primarily arising from the approximations made in Eqs.~(\ref{eq:VC-VP}) and (\ref{eq:VI-VP}). Furthermore, the relative deviation between $\sigma_\mathrm{KF-VP}^{\mu^+\mu^-}$ and $\sigma_\mathrm{Approx-VP}^{\mu^+\mu^-}$ remains about 0.3\% at $\phi=90^\circ$, and 3\% at $\phi=0^\circ$ and $180^\circ$. It is evident that the precision of $\sigma_\mathrm{Approx-VP}^{\mu^+\mu^-}$ mainly inherits from the one of $\sigma_\mathrm{Approx}^{\mu^+\mu^-}$.

\section{c.m. energy spread effect}\label{beam energy spread effect}
For the narrow resonances, such as $J/\psi$ whose decay width is 92.6 keV, the c.m. energy spreads of \(e^+e^-\) are much larger than the resonance widths. For example, the c.m. energy spread around \(J/\psi\) in BEPCII is less than 1 MeV~\cite{YaDi2019}. Therefore, the effect for the c.m. energy spread must be considered in the experimentally observed cross section $\sigma_\mathrm{exp}^{f}(W)$ by a Gaussian convolution with $\sigma_\mathrm{ISR-VP}^f(W)$ which is
\begin{eqnarray}
\sigma_\mathrm{ISR-exp}^{f}(W)&=&\int_{W-n S_E}^{W+n S_E} \frac{1}{\sqrt{2 \pi} S_E} \exp \left(\frac{-\left(W-W^{\prime}\right)^2}{2 S_E^2}\right) \nonumber
\\&&\times\sigma_\mathrm{ISR-VP}^{f}\left(W^{\prime}\right) d W^{\prime}.
\end{eqnarray}

\begin{figure*}[!htp]
    \centering
    \includegraphics[width=17.2cm]{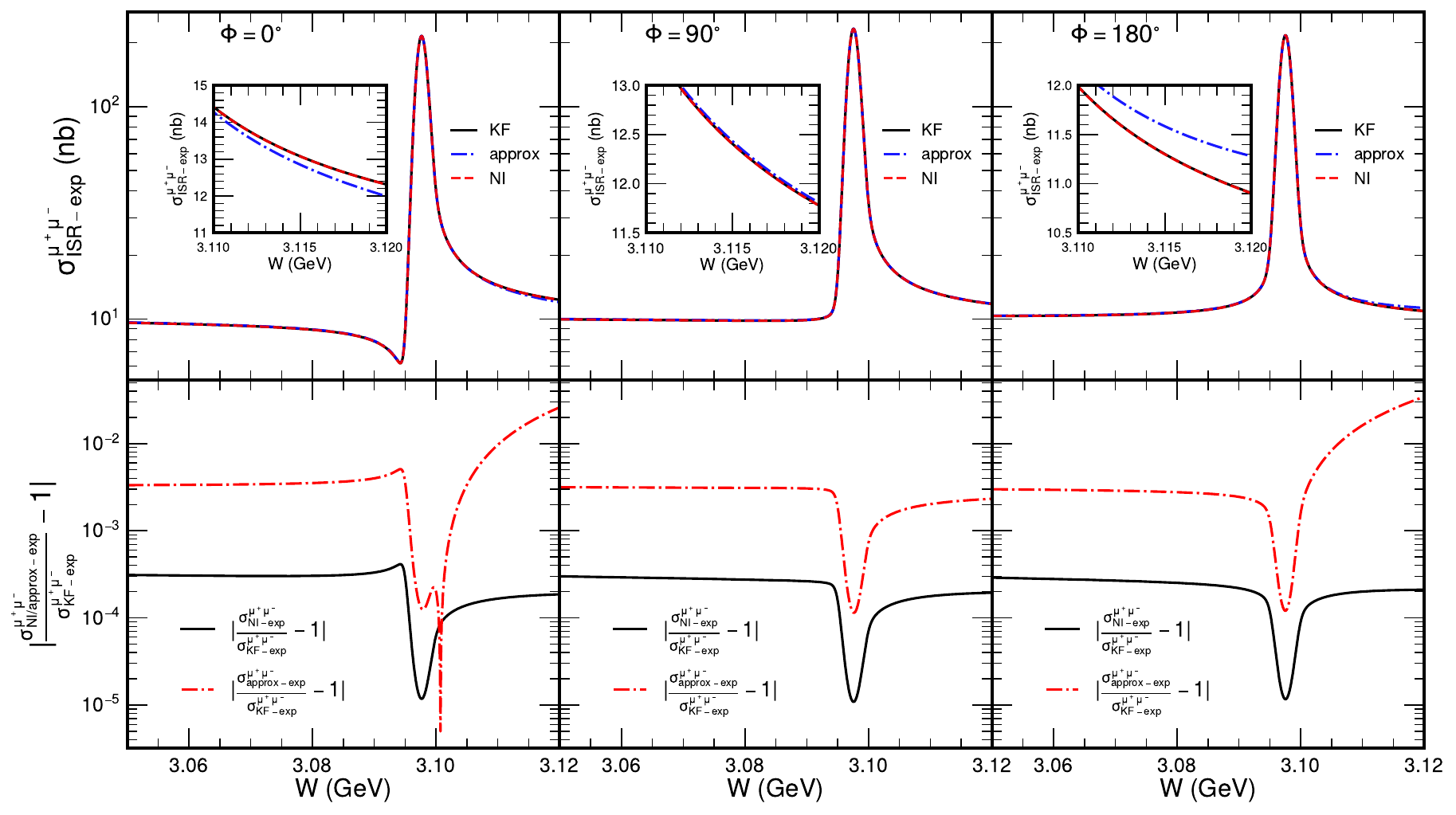}
        \captionsetup{
   justification=raggedright,
   singlelinecheck=false,
   labelsep=period
}
    \caption{\label{fig:EXP}The comparisons of ISR corrected cross section with VP effect and c.m. energy spread effect for $e^+e^-\to\mu^+\mu^-$ at $\phi=0^{\circ}$ at $\phi=0^{\circ}$, $90^{\circ}$ and $180^{\circ}$. The Born cross section parameters are set to be PDG world averages~\cite{PDG} and $W_\mathrm{min}=2.0$ GeV. In the top row, black lines represent KF method results, blue dashed lines represent approximated analytic form results and red dotted lines represent NI method results. In the bottom row, red dashed lines represent the results of $\left|\frac{\sigma^{\mu^+\mu^-}_{\text{Approx-exp}}}{\sigma^{\mu^+\mu^-}_{\text{KF-exp}}}-1\right|$ and  black lines represent the results of $\left|\frac{\sigma^{\mu^+\mu^-}_{\text{NI-exp}}}{\sigma^{\mu^+\mu^-}_{\text{KF-exp}}}-1\right|$.}
\end{figure*}

In Fig.~\ref{fig:EXP}, we compare ISR corrected cross sections including both the VP and c.m. energy spread effects $\sigma^{\mu^+\mu^-}_\mathrm{NI-exp}$ obtained through the NI  with $\sigma^{\mu^+\mu^-}_\mathrm{KF-exp}$ obtained using the KF analytic form and $\sigma^{\mu^+\mu^-}_\mathrm{Approx-exp}$ obtained using the approximated analytic form. The relative deviation between $\sigma^{\mu^+\mu^-}_\mathrm{NI-exp}$ and $\sigma^{\mu^+\mu^-}_\mathrm{KF-exp}$ remains at the order of $10^{-4}$. Moreover, the relative deviation between $\sigma^{\mu^+\mu^-}_\mathrm{KF-exp}$ and $\sigma^{\mu^+\mu^-}_\mathrm{Approx-exp}$ is still 0.3\% at $\Phi=90^\circ$ and 3\% at $\Phi=0^\circ$ and $\Phi=180^\circ$. It is evident that the Gaussian convolution of the c.m. energy spread does not impact the precision. 

\section{Fit results using toy MC samples}\label{Fit results for the toy MC cross section data}
We use the toy MC samples for $e^+e^-\to\mu^+\mu^-$ and $e^+e^-\to 2(\pi^+\pi^-)\pi^0$ around $J/\psi$ to test the KF analytic form and compare it to the approximated one. Firstly, the toy MC samples for the ISR corrected cross sections of $e^+e^-\to\mu^+\mu^-$ are generated with 1\% uncertainty at 20 c.m. energies around $J/\psi$. The minimized $\chi^2_{\mu^+\mu^-}$ is performed with the free parameters $M$, $S_E$ and $\phi$. The $\chi^2_{\mu^+\mu^-}$ is expressed as 
\begin{widetext}
\begin{equation}\label{eq:chi2mumu}
\chi^2_{\mu^+\mu^-}(M, S_E, \phi)=\sum_{i=1}^{20} \left[\frac{\sigma^{\mu^+\mu^-}_i-\sigma_{\text{ISR-exp}}^{\mu^+\mu^-}(W_i; M, S_E, \phi)}{\Delta\sigma^{\text{MC}}_i}\right]^2,
\end{equation}
\end{widetext}
where $\sigma^{\mu^+\mu^-}_i$ is the cross section at every energy point $W_i$ and $\Delta\sigma^{\mu^+\mu^-}_i$ is the corresponding uncertainty. The estimators $\widehat{M}$ and $\widehat{S_E}$ and their corresponding uncertainties $\sigma_{\widehat{M}}$ and $\sigma_{\widehat{S_E}}$ are taken as the information of the $e^+e^-$ collider. 

Secondly, the toy MC samples for the ISR corrected cross sections of $e^+e^-\to 2(\pi^+\pi^-)\pi^0$ are then generated with 1\% uncertainty at the same 20 c.m. energies around $J/\psi$. The minimized $\chi^2_{5\pi}$ is performed with the free parameters $\mathcal{A}$, $\mathcal{C}_2$ and $\Phi$, and the nuisance parameters $M$ and $S_E$. Therefore, the estimators $\widehat{\mathcal{A}}$, ${\widehat{\mathcal{C}}}_2$ and $\widehat{\Phi}$ can be obtained. The $\chi^2_{5\pi}$ is expressed as
\begin{widetext}
\begin{eqnarray}\label{eq:chi25pi}
\chi^2_{5\pi}(\mathcal{A}, \mathcal{C}_2, \Phi; M, S_E) &=& \sum_{i=1}^{20} \left[\frac{\sigma^{5\pi}_i-\sigma_{\text{ISR-exp}}^{5\pi}(W_i; \mathcal{A}, \mathcal{C}_2, \Phi, M, S_E)}{\Delta\sigma^{5\pi}_i}\right]^2\nonumber\\
&&+\left(\frac{M-\widehat{M}}{\sigma_{\widehat{M}}}\right)^2+\left(\frac{S_E-\widehat{S_E}}{\sigma_{\widehat{S_E}}}\right)^2.
\end{eqnarray}
\end{widetext}
The branching fraction for $J/\psi\to 2(\pi^+\pi^-)\pi^0$ satisfies~\cite{YaDi2019}
\begin{equation}\label{eq:Br}
\mathrm{Br} = \left(\frac{\mathcal{A}}{W^2}\right)^2\frac{\Gamma_{\mu\mu}}{\Gamma}\left|1+\mathcal{C}_2e^{i\Phi}\right|^2.
\end{equation}

\begin{figure*}[!htp]
    \centering
    \includegraphics[width=17.2cm]{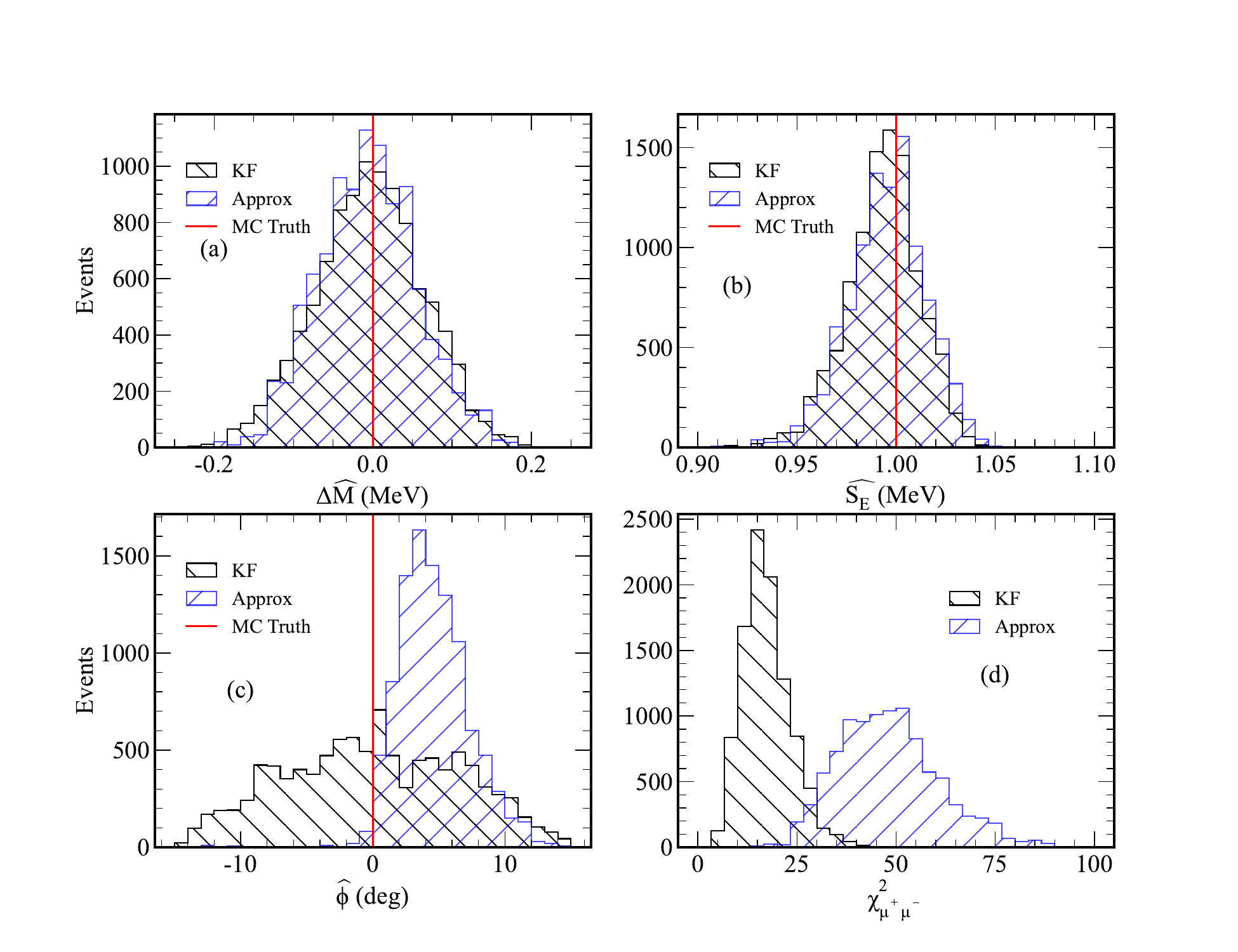}
        \captionsetup{
   justification=raggedright,
   singlelinecheck=false,
   labelsep=period
}
    \caption{\label{fig:fitmumu} The histograms of the estimators obtained from toy MC samples for $e^+e^-\to \mu^+\mu^-$. (a), (b), (c) and (d) show the distributions of $\Delta{\widehat{M}}=\widehat{M}-M^\mathrm{MCTruth}_{J/\psi}$, $\widehat{S_E}$, $\widehat\phi$ and $\chi^2_{\mu^+\mu^-}$, respectively. Histograms with black edges represent the KF analytic form, blue edges represent approximated analytic form and red lines are the values of MC truth.}
\end{figure*}

Finally, the above toy MC sample generations and $\chi^2$ minimizations are repeated 10,000 times. The fit results of $e^+e^-\to\mu^+\mu^-$ are shown in Fig.~\ref{fig:fitmumu}. The estimators of $\widehat{M}$ and $\widehat{S_E}$ are centered around the MC truth values for both KF and approximated analytic forms. Only the estimator $\widehat\phi$ of the KF analytic form is centered around the MC truth value, while the distribution of the estimator $\widehat\phi$ of the approximated analytic form deviates more than 2$\sigma$ from the MC truth value. The $\chi^2_{\mu^+\mu^-}$ of the KF analytic form is better than that of the approximated analytic form.

\begin{figure*}[!htp]
    \centering
    \includegraphics[width=17.2cm]{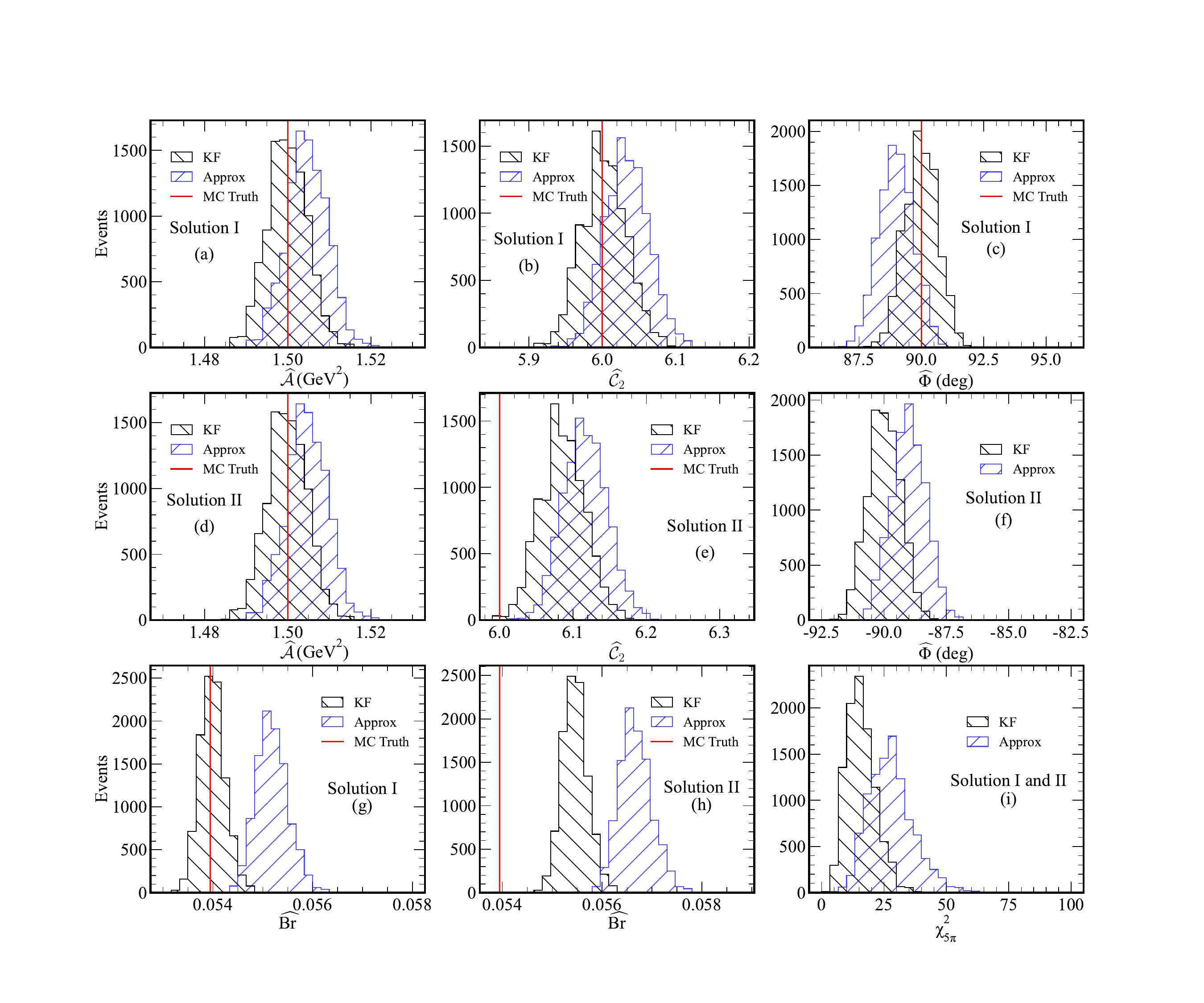}
        \captionsetup{
   justification=raggedright,
   singlelinecheck=false,
   labelsep=period
}
    \caption{\label{fig:fit5pi}   The histograms of the estimators obtained from toy MC samples for $e^+e^-\to 2(\pi^+\pi^-)\pi^0$. (a), (b), (c) and (g) show the distributions of $\widehat{\mathcal{A}}$, $\widehat{\mathcal{C}}_2$, $\Phi$ and $\widehat{\mathrm{Br}}$ for Solution I, respectively; (d), (e), (f) and (h) show the distributions of $\mathcal{A}$, $\mathcal{C}_2$,  $\Phi$ and $\widehat{\mathrm{Br}}$ for Solution II, respectively; (i) shows $\chi^2_{5\pi}$ distributions for both Solution I and II. Histograms with black edges represent the KF analytic form, blue edges represent approximated analytic form and red lines are the values of MC truth.}
\end{figure*}

The fit results for the process $e^+e^-\to 2(\pi^+\pi^-)\pi^0$ are depicted in Fig.~\ref{fig:fit5pi}. Two solutions, labeled as Solution I and II, are obtained. The distributions of $\chi^2_{5\pi}$ are identical for both Solution I and  II. The $\chi^2_{5\pi}$ of the KF analytic form is better than that of the approximated analytic form. Only in Solution I the estimator $\widehat{\Phi}$ is centered around the MC truth value of 90$^\circ$, while in Solution II, $\widehat{\Phi}$ is centered around $-90^\circ$. Furthermore, in Solution I, only the estimator $\widehat{\mathrm{Br}}$ of the KF analytic form is centered around the MC truth value, whereas the estimator of the approximated analytic form deviates significantly. The toy MC test shows that the KF analytic form avoids a subtle yet non-negligible uncertainty that arises from employing the approximated analytic form.

Table~\ref{tab:compare} shows the computing time for the fitting processes. The fitting processes took less than 10 seconds for $e^+e^-\to\mu^+\mu^-$ and approximately 6 seconds for $J/\psi\to2(\pi^+\pi^-)\pi^0$ with the KF analytic form, while it required about 20 and 16 minutes with NI. Consequently, the KF analytic form can improve the fitting speed by more than 160 times compared to NI.
\begin{table*}[!htp]
      \captionsetup{
   justification=raggedright,
   singlelinecheck=false,
   labelsep=period
}
\caption{\label{tab:compare}
The comparison of the computing times and the number of function calls (NFC) for fitting with the typical MC samples. Fitting codes are written in C++ and run on a single core of an Intel 3 GHz CPU.}
\begin{ruledtabular}
\begin{tabular}{ccccccc}
&\multicolumn{2}{c}{$e^+e^-\to\mu^+\mu^-$}&
\multicolumn{4}{c}{$e^+e^-\to5\pi$}\\
&\multicolumn{2}{c}{}&\multicolumn{2}{c}{Solution I}&
\multicolumn{2}{c}{Solution II}\\
\hline
Method&Time & NFC &Time &NFC&Time &NFC\\
KF & 9.3 s& 3600 &6.1 s & 3900 &5.9 s& 3920 \\
Approx &0.98 s & 3160 &3 min 53 s & 4360 &4 min 11 s & 4740\\
NI & 29 min 27 s& 4500 & 16 min 20 s & 3960 &16 min 5 s & 3960\\
\end{tabular}
\end{ruledtabular}
\end{table*}

\begin{figure*}[!htp]
    \centering
    \includegraphics[width=17.2cm]{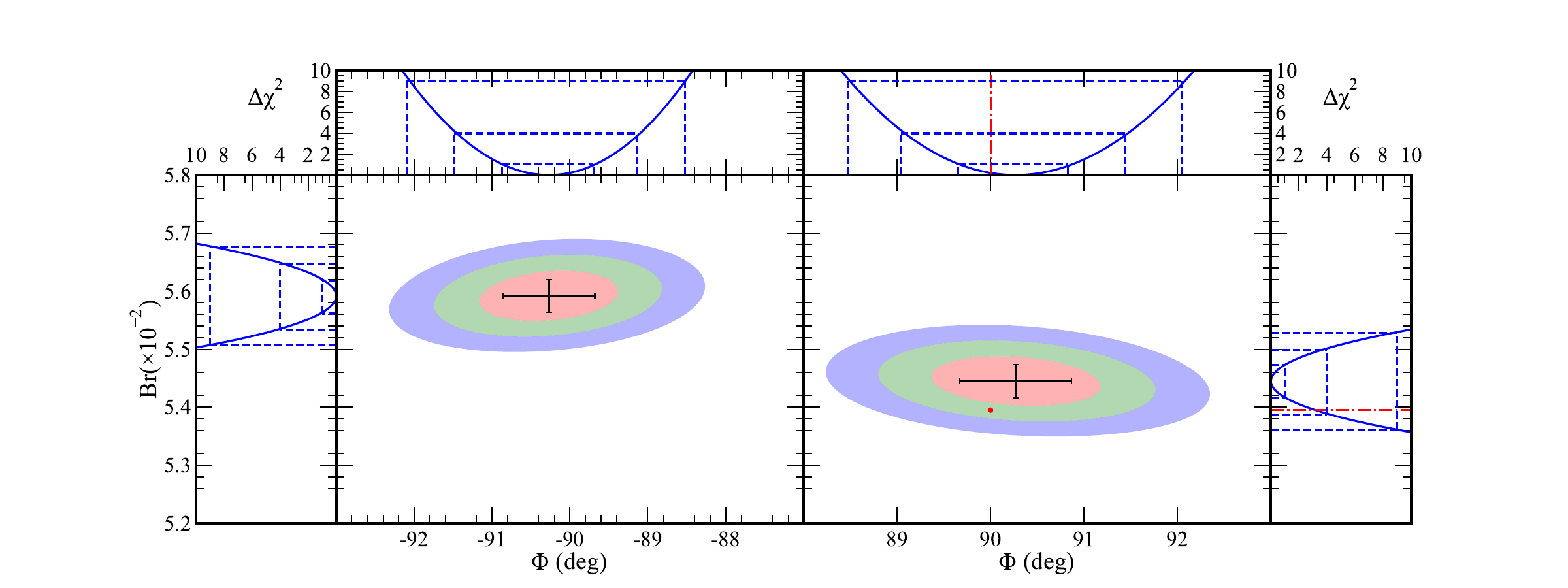}
        \captionsetup{
   justification=raggedright,
   singlelinecheck=false,
   labelsep=period
}
    \caption{\label{fig:contour}   Confidence regions of $\Phi$ and branching fraction $\mathrm{Br}$ from the fit of one typical toy MC sample set for $e^+e^-\to2(\pi^+\pi^-)\pi^0$. The $1\sigma$, $2\sigma$ and $3\sigma$ two-dimensional confidence regions are estimated using $\Delta \chi^2$ values of 2.30 (red), 6.18 (green) and 11.83 (blue) relative to the best fit. The left and right upper panels provide the one-dimensional $\Delta \chi^2$ for $\Phi$ obtained by profiling branching fraction (blue line), the dashed lines mark the corresponding $1\sigma$, $2\sigma$, and $3\sigma$ intervals; and the red dotted line represents MC truth value. The left and right panels are the same, but for branching fraction, with $\Phi$ profiled. The black points mark the best estimates, and the error bars display their one-dimensional $1\sigma$ confidence intervals. The red dot and dash-dotted lines mark the MC truth values. All the fit results took 55 minutes.}
\end{figure*}

As it is shown in Fig.~\ref{fig:observed}, the relative derivation between the analytic result and the numerical integration can be pushed down to the level of $10^{-12}$. Moreover, the different contributions, such as VP effect, can be switched off and on in the case of numerical integration and then test physical effects. As it is shown in Table \ref{tab:compare}, for one typical toy MC sample set, the time consumption of the fitting for $J/\psi\to2(\pi^+\pi^-)\pi^0$ is only about 16 minutes. It seems that there is no real practical reason to avoid numerical integration, as it is reliable and fast enough with modern computers. However, the current data analysis no longer solely focuses on reporting the confidence region of the branching fraction $\mathrm{Br}$ or the phase angle $\Phi$ separately. Equation~\eqref{eq:Br} shows that the branching fraction for $J/\psi\to2(\pi^+\pi^-)\pi^0$ is correlated to the phase angle $\Phi$. It is better to provide the confidence regions of both branching fraction $\mathrm{Br}$ and phase angle $\Phi$. The 1$\sigma$, 2$\sigma$, and 3$\sigma$ two-dimensional confidence regions are estimated using the $\Delta\chi^2$ value of 2.3, 6.18 and 11.83 relative to the best fit, respectively, as shown in Fig.~\ref{fig:contour}. The computation time for obtaining the two-dimensional confidence regions is only 55 minutes using the KF analytic form. Given that the KF analytic form is over 160 times faster than the NI method with similar NFC, the estimated computation time with the NI method exceeds 6 days. Consequently, the impracticality of the NI method is evident when considering the computation time for two-dimensional confidence regions.

\section{Discussions and Conclusions}\label{Discussions and Conclusions}
The ISR correction is essential for precise cross section measurements in $e^+e^-$ annihilation. For the first time, we present the exact analytic forms for the ISR corrected cross sections around a narrow resonance with the QED structure function formalism by KF. The KF analytic forms can accelerate the regression procedure used to extract physical parameters, such as the branching fraction, by over 160 times. Utilizing the toy MC samples, we reveal a non-negligible few percent systematic uncertainty caused by the approximated analytic form. 

The KF formalism is well known~\cite{E1985}. It has been applied for description of QED radiative corrections to many observables in high-energy physics, including cross sections of $e^+e^-$ annihilation processes at different energies. Moreover, the KF formalism was realized in several MC codes. In particular, BabaYaga~\cite{Giovanni2006} and MCGPJ~\cite{Arbuzov2006} programs are widely used at $e^+e^-$ colliders. It is also coded within the framework of BesEvtGen for the BESIII experiment~\cite{Ping2014}. Since there is a certain freedom in the exponentiation prescription of the QED radiative corrections~\cite{Jadach1991}, there are many other exponentiation recipes of the QED structure function formalism~\cite{Montagna1997}. Although the uncertainty caused by VP and c.m. energy spread effects is below 0.1\%, it does not mean that the theoretical uncertainty of the KF approach can reach the same order.  As one can learn from Refs.~\cite{Giovanni2006,Arbuzov2006,Actis2010}, such a high precision cannot be provided without application of the complete one-loop corrections and the leading effects in orders higher than $O(\alpha^2)$~\cite{Blumlein:2011mi, Blumlein:2020jrf}. Moreover, certain observables, especially around resonances, require calculation of next-to-leading logarithmic corrections of the order $O(\alpha^2L^1)$ which appear beyond the original formalism of Kuraev and Fadin. Comparisons with the results of alternative realizations, e.g., with BabaYaga~\cite{Giovanni2006} and MCGPJ~\cite{Arbuzov2006}, should be performed to estimate the resulting theoretical precision.  The difference should be taken as the systematic uncertainties of event selection efficiencies for the experimentally observed cross section data.

For the currently running $e^+e^-$ collision experiments,  the BESIII experiment has accumulated 10 billion $J/\psi$ events~\cite{Ablikim2021} and 3 billion $\psi(3686)$ events~\cite{Ablikim2024} which are at least 6 times larger than those used in previous BESIII measurements. Furthermore, the Belle II experiment plans to take about 500 fb$^{-1}$ data for each vector bottomonium state~\cite{Kou2020} which are at least ten or hundreds of times larger than the Belle experiment. The KF analytic forms will be helpful to handle the ISR correction properly in currently running experiments at BESIII and Belle II, as well as in planned ones, such as the super-tau-charm factories~\cite{Achasov2024} and the super-$J/\psi$ factory~\cite{Yuan2021}.

\begin{acknowledgments}
X. Z. would like to acknowledge useful conversations with Y. D. Wang and K. Zhu. This work has been supported by the National Natural Science Foundation of China (NSFC) under Contract No. 12192265 and Joint Fund of Research utilizing Large-Scale Scientific Facility of the NSFC and CAS under Contract No. U2032114.
\end{acknowledgments}

\appendix
\section{ANALYTIC FORMS FOR BORN CROSS SECTIONS}\label{Appendix A} 

The Born cross section of the process $e^+e^-\to\mu^+\mu^-$ is~\cite{YaDi2019}
\begin{eqnarray}
    \sigma_\mathrm{Born}^{\mu^+\mu^-}(W)&=&\frac{4 \pi \alpha^2}{3 W^2}\left|1+\frac{W^2}{M}
    \nonumber\right.\\&&\left.\times \frac{3 \sqrt{\Gamma_{e e} \Gamma_{\mu \mu}}}{\alpha\left(W^2-M^2+i M \Gamma\right)} e^{i \phi}\right|^2, 
\end{eqnarray}
which can be taken apart into three parts

\begin{equation}
    \sigma^{\mu^+\mu^-}_{B_c}(W) = \frac{4\pi\alpha^2}{3W^2},
\end{equation}
\begin{equation}
    \sigma^{\mu^+\mu^-}_{B_R}(W) = \frac{12\pi W^2\Gamma_{ee}\Gamma_{\mu\mu}}{M^2((W^2-M^2)^2+M^2\Gamma^2)},
\end{equation}
and
\begin{eqnarray}
    \sigma^{\mu^+\mu^-}_{B_I}(W) &=&  \frac{8\pi \alpha \sqrt{\Gamma_{ee}\Gamma_{\mu\mu}}}{M((W^2-M^2)^2+M^2\Gamma^2)}\nonumber\\&&\times\left((W^2-M^2)\text{cos}\phi + M\Gamma\text{sin}\phi \right).
\end{eqnarray}

The Born cross section of the process $e^+e^-\to2(\pi^+\pi^-)\pi^0$ or $e^+e^-\to\eta\pi^+\pi^-$ with $\eta\to\pi^+\pi^-\pi^0$, abbreviated as $5\pi$ is~\cite{YaDi2019}
\begin{eqnarray}
\sigma_\mathrm{Born}^{5\pi}(W)  &=&\left(\frac{\mathcal{A}}{W^2}\right)^2 \frac{4 \pi \alpha^2}{3W^2} \left|1+\frac{W^2}{M}\right.\nonumber\\&&\left.\times\frac{3 \sqrt{\Gamma_{e e} \Gamma_{\mu \mu}} \mathcal{C}_1 e^{i \phi}\left(1+\mathcal{C}_2 e^{i \Phi}\right)}{\alpha \left(W^2-M^2+i M \Gamma\right)}\right|^2,
\end{eqnarray}
which can also be taken apart into three parts
\begin{equation}
    \sigma_{B_c}^{5\pi}(W) = \left(\frac{\mathcal{A}}{W^2}\right)^2 \frac{4 \pi \alpha^2}{3W^2},
\end{equation}
\begin{eqnarray}
    \sigma_{B_R}^{5\pi}(W) &=& \mathcal{C}_1^2(1+\mathcal{C}_2^2+2\mathcal{C}_2\text{cos}\Phi)\nonumber\\&&\times\frac{12\pi\mathcal{A}^2\Gamma_{ee}\Gamma_{\mu\mu}}{M^2W^2((W^2-M^2)^2+M^2\Gamma^2)},
\end{eqnarray}
and
\begin{eqnarray}
    \sigma_{B_I}^{5\pi}(W) &=& \frac{8\pi\alpha\mathcal{A}^2\sqrt{\Gamma_{ee}\Gamma_{\mu\mu}}}{W^4M((W^2-M^2)^2+M^2\Gamma^2)}\nonumber\\
    &&\times\left[(W^2-M^2) \mathcal{C}_1(\text{cos}\phi + \mathcal{C}_2\text{cos}\phi\text{cos}\Phi\nonumber\right.\\&&\left.-\mathcal{C}_2\text{sin}\phi\text{sin}\Phi)+M\Gamma\mathcal{C}_1(\text{sin}\phi + \nonumber\right.\\&&\left.\mathcal{C}_2\text{cos}\phi\text{sin}\Phi+\mathcal{C}_2\text{sin}\phi\text{cos}\Phi)\right].
\end{eqnarray}

\section{ANALYTIC FORMS OF ISR CORRECTED CROSS SECTIONS}\label{AppendixB}
The continuum part of the ISR corrected cross section for $e^+e^-\to\mu^+\mu^-$ is
\begin{eqnarray}
    \sigma_C^{\mu^+\mu^-} &=& \frac{4\pi\alpha^2}{3W^2}I_0,
\end{eqnarray}
the resonance part is
\begin{eqnarray}
    \sigma^{\mu^+\mu^-}_R &=& \frac{6\pi\Gamma_{ee}\Gamma_{\mu\mu}W^2}{M^5\Gamma}\left[\left(1+\frac{A}{B}\right)I_1-\frac{1}{B}I_2\right]\nonumber\\&&+\mathrm{c.c.},
\end{eqnarray}
and the interference part is
\begin{eqnarray}
    \sigma^{\mu^+\mu^-}_I &=& \left[\frac{4\pi\alpha\sqrt{\Gamma_{ee}\Gamma_{\mu\mu}}}{M^3\Gamma}(\Gamma\text{sin}\phi-M\text{cos}\phi)+\nonumber\right.\\&&\left.\frac{4\pi\alpha\sqrt{\Gamma_{ee}\Gamma_{\mu\mu}}}{M^4\Gamma}W^2\text{cos}\phi\left(1+\frac{A}{B}\right)\right]I_1\nonumber\\&&~~ 
-\frac{I_2W^2}{B}\frac{4\pi\alpha\sqrt{\Gamma_{ee}\Gamma_{\mu\mu}}}{M^4\Gamma}\text{cos}\phi  + \mathrm{c.c.},
\end{eqnarray}
where $\mathrm{c.c.}$ represents the complex conjugate of the former part.

The continuum part of the ISR corrected cross section for $e^+e^-\to 5\pi$ is
\begin{eqnarray}
     \sigma^{5\pi}_C &=& \frac{4\pi\alpha^2\mathcal{A}^2}{3W^6}I_3,
\end{eqnarray}
the resonance part is
\begin{eqnarray}
    \sigma^{5\pi}_R &=& \mathcal{C}_1^2(1+\mathcal{C}_2^2+2\mathcal{C}_2\text{cos}\Phi)\frac{6\pi\mathcal{A}^2\Gamma_{ee}\Gamma_{\mu\mu}}{\Gamma M^5W^2}\nonumber\\&&\left(\frac{1}{A+B}I_0+\frac{B}{A+B}I_1  \right)+\mathrm{c.c.},
\end{eqnarray}
and the interference part is
\begin{eqnarray}
    \sigma^{5\pi}_I &=&\frac{4\pi\alpha\mathcal{A}^2\sqrt{\Gamma_{ee}\Gamma_{\mu\mu}}}{M}\left\{\left[\frac{\mathcal{C}_1}{ W^4M^2}\nonumber\right.\right.\\&&\left.\left.\times(\text{sin}\phi + \mathcal{C}_2\text{cos}\phi\text{sin}\Phi+\mathcal{C}_2\text{sin}\phi\text{cos}\Phi)\right.\right.\nonumber\\ 
    &&\left.\left.-\frac{\mathcal{C}_1(\text{cos}\phi + \mathcal{C}_2\text{cos}\phi\text{cos}\Phi-\mathcal{C}_2\text{sin}\phi\text{sin}\Phi)}{\Gamma W^4M}\right]\nonumber\right.\\&&\left.\times\left[ \frac{B}{(A+B)^2}I_0+ \left(\frac{B}{A+B}\right)^2I_1+\frac{1}{A+B}I_4
    \right]+\nonumber\right.\\&&\left.\frac{\mathcal{C}_1(\text{cos}\phi + \mathcal{C}_2\text{cos}\phi\text{cos}\Phi-\mathcal{C}_2\text{sin}\phi\text{sin}\Phi)}{\Gamma W^2M^3}\left( \frac{1}{A+B}I_0  
    \right.\right.\nonumber\\ 
    &&\left.\left.+ \frac{B}{A+B}I_1\right)\right\}+\mathrm{c.c.},
\end{eqnarray}
where  $ A=\frac{\Gamma}{M}+i\left[\left(\frac{W}{M}\right)^2-1\right]$, $B = -i\left(\frac{W}{M}\right)^2$ and $b = 1-\left(\frac{W_{\text{min}}}{W}\right)^2$. 

The five integrations $I_0$, $I_1$, $I_2$, $I_3$ and $I_4$ are
\begin{widetext}
\begin{eqnarray}\label{I0}
    I_0 &=& \beta(1+\delta)\mathcal{B}(b;\beta,0) - \frac{b\beta}{2} + \log(1-b)\left(\frac{1}{4}\beta^2+\frac{\beta}{2}+\frac{3}{8}\beta^2b\right) + \frac{1}{16}\beta^2\log^2(1-b)\nonumber\\
&&~~-\frac{\beta^2}{2}b\log b-\frac{\beta^2}{2}\left(\text{Li}_2(1-b)-\frac{\pi^2}{6}-\text{Li}_2(b)\right),
\end{eqnarray}
\begin{eqnarray}\label{I1}
    I_1 &=& \beta(1+\delta)\left(\frac{b^{\beta}{}\,_2F_1(1,\beta;1+\beta;-bB/A)}{A\beta}\right) + \left(\frac{\beta^2}{8}+\frac{\beta}{2}\right)\left(\frac{b}{B}-\frac{A}{B^2}\log\frac{A+Bb}{A}\right) - \nonumber\\&&~~\left(\beta+\frac{3}{4}\beta^2\right)\frac{1}{B}\log\frac{A+Bb}{A}
 + \frac{3}{4}\beta^2\left[-\frac{1}{B}\left[\text{Li}_2\left(\frac{A+Bb}{A+B}\right)-\text{Li}_2\left(\frac{A}{A+B}\right)
 \right.\right.\nonumber\\&&~~\left.\left.+\log\frac{B}{A+B}\log\frac{A+Bb}{A}\right]\right] - \frac{3}{8}\beta^2\frac{1}{B}\left[-b+(b-1)\log(1-b)+
 \right.\nonumber\\&&~~\left.\frac{A}{B}\left[\text{Li}_2\left(\frac{A+Bb}{A+B}\right)-\text{Li}_2\left(\frac{A}{A+B}\right)+\log\frac{B}{A+B}\log\frac{A+Bb}{A}\right]\right]
\nonumber\\&&~~ -\beta^2\frac{1}{B}\left[-\frac{\pi^2}{6}+\frac{1}{2}\log^2\left(1+\frac{B}{A}b\right)+\text{Li}_2\left(\frac{A}{A+Bb}\right)-\log\frac{B}{A}\log\left(1+\frac{B}{A}b\right)\right]
\nonumber\\&&~~+\frac{1}{2}\beta^2\frac{1}{B}\left[b(\log b-1) - \frac{A}{B}\left[-\frac{\pi^2}{6}+\frac{1}{2}\log^2\left(1+\frac{B}{A}b\right)+\text{Li}_2\left(\frac{A}{A+Bb}\right)-
 \right.\right.\nonumber\\&&~~\left.\left.\log\frac{B}{A}\log\left(1+\frac{B}{A}b\right)\right]\right] -\frac{1}{2}\beta^2\frac{1}{A} \left[-\text{Li}_2(b) + \left[\text{Li}_2\left(\frac{A+Bb}{A+B}\right)- \right.\right.\nonumber\\&&~~\left.\left.\text{Li}_2\left(\frac{A}{A+B}\right)+\log\frac{B}{A+B}\log\frac{A+Bb}{A}\right]\right],
\end{eqnarray}
\begin{eqnarray}\label{I2}
    I_2 &=& \beta(1+\delta)\frac{b^{\beta}}{\beta}+\frac{\beta^2}{2}\text{Li}_2(b) + b^2\left(\frac{\beta}{4}+\frac{1}{32}\beta^2\right) + b\left(-\beta-\frac{5}{16}\beta^2\right)\nonumber\\
&&~~+\log(1-b)\left(-\frac{9}{16}\beta^2+\frac{3}{4}\beta^2b-\frac{3}{16}\beta^2b^2\right)+\log b\left(\frac{\beta^2b^2}{4}-\beta^2b\right),
\end{eqnarray}
\begin{eqnarray}\label{I3}
    I_3 &=& \beta(1+\delta)\mathcal{B}(b;\beta,-2) + \left(\frac{\beta^2}{8}+\frac{\beta}{2}\right)\frac{-1+(1-b)^3+b(b+3-3b)}{2(1-b)^{3}}\nonumber\\
&&~~-\frac{1}{2}\left(\beta+\frac{3}{4}\beta^2\right)\left(\frac{1}{(1-b)^2}-1\right)-\frac{\beta^2}{8} \frac{1-(1-b)^2+2\log(1-b)}{4(1-b)^2}  \nonumber\\
&&~~-\frac{3}{4}\beta^2\log(1-b) -\frac{\beta^2}{4}  \frac{b(b-1-(b-2)\log b)}{(1-b)^2}+\frac{\beta^2}{2}\frac{b\log b}{b-1}\nonumber\\
&&~~-\frac{1}{2}\beta^2\left(-\text{Li}_2(b)-\frac{1}{2}\log^2(1-b)\right)-\frac{\beta}{8}\frac{b+\text{log}(1-b)}{1-b} ,
\end{eqnarray}
and
\begin{eqnarray}\label{I4}
    I_4 &=& \beta(1+\delta)\mathcal{B}(b;\beta,-1) + \frac{\beta^2}{2}(\text{Li}_2(b)-\text{Li}_2(1-b)+\frac{\pi^2}{6}) + \frac{b}{1-b}\left(-\frac{3}{4}\beta^2-\frac{\beta}{2}\right) \nonumber\\
&&~~ +\log(1-b)\left(\frac{\beta}{2}-\frac{3}{8}\beta^2\right)-\frac{b\log b}{1-b}\frac{\beta^2}{2}+\log^2(1-b)\frac{\beta^2}{16}-\frac{\beta^2}{8}\frac{\log(1-b)}{1-b}.
\end{eqnarray}
\end{widetext}
The special functions used in the above formulas are
\begin{equation}
    \text{Li}_2(x) \equiv -\int_0^x\frac{\log(1-u)}{u}\text{d}u,
\end{equation}
\begin{equation}
    \mathcal{B}(x;a,b) \equiv \int_0^x t^{a-1}(1-t)^{b-1}\text{d}t,
\end{equation}
and ${}\,_2F_1(a,b;c;x)$, where $\text{Li}_2(x)$ is the Spence's function, $\mathcal{B}(x;a,b)$ is the incomplete beta function and ${}_2F_1(a,b;c;x)$ is the Gaussian hypergeometric function~\cite{Zhang1996}.

\nocite{*}
\bibliography{ISR}

\end{document}